# Hydrogen site-dependent physical properties of hydrous magnesium silicates: implications for water storage and transport in the mantle transition zone

Zifan Wang[1], Yu He[1,2], Ho-kwang Mao[1,3], Duck Young Kim[1]✉

[1]Center for High Pressure Science & Technology Advanced Research (HPSTAR), Shanghai 201203, P.R. China

[2]Key Laboratory of High-Temperature and High-Pressure Study of the Earth's Interior, Institute of Geochemistry, Chinese Academy of Sciences, Guiyang 550081, Guizhou, China

[3] Shanghai Key Laboratory of Material Frontiers Research in Extreme Environments (MFree), Institute for Shanghai Advanced Research in Physical Sciences (SHARPS), Shanghai 201203, P.R. China

✉e-mail: duckyoung.kim@hpstar.ac.cn

**Abstract**

The Earth's mantle transition zone (MTZ) is widely recognized as a major water reservoir, exerting significant influence on the planet's water budget and deep cycling processes. Here, we employ crystal structure prediction and first-principles calculations to identify a series of stable hydrous magnesium silicate phases under transition zone conditions. Our results reveal a pressure-induced hydrogen substitution mechanism in wadsleyite, where $H^+$ preferentially migrates from $Mg^{2+}$ sites to $Si^{4+}$ sites near ~410 km depth. This transformation leads to a substantial decrease in electrical conductivity, consistent with geophysical observations. We estimate the water content in the MTZ to be approximately **1.6 wt%**, aligning with seismic and conductivity constraints. Furthermore, using machine learning-enhanced molecular dynamics, we discover *double superionicity* in hydrous wadsleyite and ringwoodite at temperatures exceeding 2000 K, wherein both $H^+$ and $Mg^{2+}$ exhibit high ionic mobility. This dual-ion superionic state has potentially profound implications for mass transport, electrical conductivity, and magnetic dynamo generation in rocky super-Earth exoplanets.

**Introduction**

Earth, often referred to as the "water planet", has over 70% of its surface covered by liquid water. Water is transported into Earth's interior through the subduction process, playing a crucial role in the planet's evolution and dynamics. Growing evidence suggests that significant amounts of water are stored not only on Earth's surface but also deep within its interior[1-4]. Specifically, subduction processes carry water-bearing sediments and minerals into the mantle. Some of this water returns to the surface through mantle degassing, such as volcanic activity beneath mid-ocean ridges and oceanic islands. However, the amount of water degassed is less than the water subducted into the mantle, suggesting the existence of hidden water reservoirs within Earth's interior[5,6].

This internal water is stored in numerous hydrous minerals. Various hydrous magnesium silicates, such as phase A ($Mg_7Si_2O_{14}H_6$), phase E ($Mg_2SiO_6H_4$), superhydrous phase B ($Mg_{10}Si_3O_{18}H_4$), phase D ($MgSi_2O_6H_2$), and phase H ($MgSiO_4H_2$), have been identified as prime water carriers in the upper and possibly lower mantle[7-11]. Additionally, water can be stored in nominally anhydrous minerals (NAMs), including forsterite (α phase), wadsleyite (β phase), ringwoodite (γ phase), and bridgmanite. Studies have shown that wadsleyite and ringwoodite can hold up to ~3.3 wt % and 1.3 wt % water, respectively, indicating that MTZ could serve as a significant water reservoir[13,14].

Hydrous minerals within the mantle are essential for water transport and cycling inside Earth. Water incorporation in these minerals influences their physical properties, including electrical conductivity[13,15,16], thermal conductivity[17,18], sound velocity[19,20], and viscosity[21]. Water content also affects phase boundaries and phase transitions, such as shifting the α-β transition boundary to lower pressures and raising the postspinel transition pressure[22,23].

Recent research has utilized various analytical techniques, including IR spectroscopy[24-26], Raman spectroscopy[27-29], NMR spectroscopy[30,31], and theoretical calculations[32-36], to investigate water incorporation in NAMs. However, several knowledge gaps persist. Few theoretical studies focus on the exploration of the most stable hydrous structures in major NAMs, which may overlook critical stable configurations. Additionally, there remains debate surrounding the actual water content in MTZ. The water contents deduced by the measured conductivities of hydrous wadsleyite and ringwoodite have differences with one order of magnitude[37,38], indicating that other important factors affecting water content may not yet be fully considered.

In this study, we performed structure predictions on a range of hydrous minerals, including forsterite (α-$Mg_2SiO_4$), wadsleyite (β-$Mg_2SiO_4$), ringwoodite (γ-$Mg_2SiO_4$), bridgmanite ($MgSiO_3$), and periclase (MgO), to identify their most stable crystal structures in hydrous form under mantle conditions. We assessed their thermal and dynamical stability by calculating formation enthalpies, Gibbs free energies, and phonon spectra. Our analysis identified three stable structures—silicon substituted forsterite (α-Si): $Mg_{32}Si_{15}O_{64}H_4$, magnesium substituted wadsleyite (β-Mg): $Mg_{31}Si_{16}O_{64}H_2$, and silicon substituted wadsleyite β-Si: $Mg_{32}Si_{15}O_{64}H_4$—and revealed a transition from β-Mg to β-Si under MTZ's conditions. Additionally, we found that electrical conductivity decreases with substitution transfer, leading to a re-evaluation of water content in the MTZ. Finally, molecular dynamics calculations using machine learning potentials uncovered double superionic states in hydrous wadsleyite and ringwoodite, suggesting these states may contribute to strong magnetic field generation in super-Earths.

**Results**

**Structure and stability**

To determine the energetically favorable hydrogen positions in defect-bearing hydrous magnesium silicates, we employed ab initio random structure searching (AIRSS)[39]. In our simulations, individual Mg or Si atoms were selectively removed from the lattice, and 2 or 4 hydrogen atoms, respectively, were randomly placed within a 1.5 Å radius of the resulting vacancy as shown in Fig. 1a. This procedure was repeated approximately 1000 times for each defect type, and the resulting structures were fully relaxed to evaluate their total energies. Additionally, we explored the possibility of defect clustering by introducing vacancies at two neighboring cation sites to assess potential hydrogen aggregation. Our results indicate that clustered hydrogen substitutions yield higher total energies compared to isolated defects, suggesting that single-site substitution is energetically more favorable. Consequently, we adopted the single-site substitution model for subsequent analyses.

We assessed the thermodynamic stability of the predicted hydrous magnesium silicates by calculating their formation enthalpies over a pressure range of 0–100 GPa. This evaluation was conducted along several plausible reaction pathways involving the incorporation of water into anhydrous mantle minerals. The reference phases used in these reactions include $Mg_2SiO_4$ (in its α-, β-, and γ-polymorphs), $MgSiO_3$, $MgO$, $SiO_2$, and $H_2O$ — all of which represent major constituents of the Earth's mantle.

$$16Mg_2SiO_4\ (\alpha) + H_2O = Mg_{31}Si_{16}O_{64}H_2\ (\alpha\text{-Mg}) + MgO \quad (1)$$

$$16Mg_2SiO_4\ (\alpha) + 2H_2O = Mg_{32}Si_{15}O_{64}H_4\ (\alpha\text{-Si}) + SiO_2 \quad (2)$$

$$16Mg_2SiO_4\ (\beta) + H_2O = Mg_{31}Si_{16}O_{64}H_2\ (\beta\text{-Mg}) + MgO \quad (3)$$

$$16Mg_2SiO_4\ (\beta) + 2H_2O = Mg_{32}Si_{15}O_{64}H_4\ (\beta\text{-Si}) + SiO_2 \quad (4)$$

$$8Mg_2SiO_4\ (\gamma) + H_2O = Mg_{15}Si_8O_{32}H_2\ (\gamma\text{-Mg}) + MgO \quad (5)$$

$$8Mg_2SiO_4\ (\gamma) + 2H_2O = Mg_{16}Si_7O_{32}H_4\ (\gamma\text{-Si}) + SiO_2 \quad (6)$$

$$16MgSiO_3\ (Prv) + H_2O = Mg_{15}Si_{16}O_{48}H_2\ (Prv\text{-Mg}) + MgO \quad (7)$$

$$16MgSiO_3\ (Prv) + 2H_2O = Mg_{16}Si_{15}O_{48}H_4\ (Prv\text{-Si}) + SiO_2 \quad (8)$$

$$31MgO\ (Per) + H_2O = Mg_{31}O_{32}H_2\ (Per\text{-Mg}) \quad (9)$$

Our calculations reveal that the formation enthalpies of three representative hydrous magnesium silicates — α-Si, β-Mg, and β-Si — are negative throughout the pressure range relevant to Earth's upper mantle and MTZ when zero-point energy (ZPE) is included. This indicates that these minerals are thermodynamically stable in the upper mantle and MTZ. Specifically, α-Si becomes stable above 10 GPa, β-Mg is stable below 12 GPa, and β-Si exhibits stability above 10 GPa. In contrast, the formation enthalpies of α-Mg and γ-Si remain relatively small but positive up to 100 GPa, implying that these phases may attain thermodynamic stability at pressures exceeding 100 GPa, potentially relevant to deep lower mantle or super-Earth's interiors. The pressure dependence of formation enthalpies, with ZPE corrections, for all studied hydrous phases across the 0–100 GPa range is presented in Fig. 1b.

The analysis of formation enthalpies reveals several noteworthy trends that shed light on the behavior of hydrous magnesium silicates under mantle conditions. First, a clear divergence emerges in the pressure dependence of substitution types: Mg-site substituted phases generally exhibit increasing formation enthalpy with pressure, whereas Si-site substituted phases show a decreasing trend. This suggests that, as pressure increases, hydrogen substitution into Si sites becomes energetically more favorable than into Mg sites. Second, none of the hydrous phases relevant to the lower mantle — including Prv-Mg, Prv-Si, and Per-Mg — are found to be thermodynamically stable up to 100 GPa. This instability indicates that major magnesium silicates in the lower mantle likely remain anhydrous, supporting the notion of a predominantly dry lower mantle. Finally, we observe a distinct crossover in the enthalpy curves of β-Mg and β-Si, both of which have negative formation enthalpies but intersect near the pressure conditions of the MTZ. This crossover marks a pressure-induced shift in the preferred hydrogen substitution site from Mg to Si, providing the theoretical evidence for hydrogen site transfer within MTZ — a mechanism that could have significant implications for hydrogen mobility, electrical conductivity, and water storage in the deep Earth.

To further evaluate the thermal stability of key hydrous phases in MTZ, we computed the Gibbs free energies of α-Si, β-Mg, and β-Si using the quasi-harmonic approximation (QHA). The results confirm that all three structures remain thermodynamically stable under relevant high-temperature conditions. Importantly, the hydrogen substitution transfer from Mg to Si sites persists at elevated temperatures, demonstrating the robustness of this phenomenon. Our calculations also reveal that the transition pressure associated with the substitution transfer is temperature-dependent, shifting by approximately 4 GPa at 1500 K (Fig. 1c). These findings provide compelling thermodynamic evidence that the substitution transfer mechanism remains active under MTZ's conditions and may play a key role in influencing the deep Earth's hydrogen distribution and related physical properties. Notably, this substitution transfer is from a low-water-content phase to a high-water-content phase, with water content increasing from ~0.8 wt% to ~1.6 wt%. The water contents were estimated by dividing the mass of incorporated $H_2O$ by the mass of per chemical formula of each hydrous mineral.

To assess the dynamic stability of the predicted hydrous magnesium silicates, we computed their phonon dispersion relations over the pressure range of 0–100 GPa. The resulting phonon spectra exhibit no imaginary frequencies across the examined pressure conditions, confirming that these phases are dynamically stable under Earth's mantle pressures. Fig. 1d summarizes the pressure stability ranges for each identified hydrous phase, illustrating their vibrational robustness across a wide span of mantle-relevant conditions. These results reinforce the structural viability of the proposed hydrous minerals throughout the upper mantle and MTZ.

To investigate the bonding nature of hydrogen, we compared the distances between hydrogen and its nearest oxygen atom (O–H) and its second-nearest oxygen atom (O···H), as illustrated in Fig. S2. At low pressure, the bonding asymmetry is most pronounced, with the largest difference between O–H and O···H distances. As pressure increases, hydrous magnesium silicates with Mg-site substitution exhibit O–H···O bond symmetrization, indicating a rapid weakening of the O–H bond. In contrast, phases with Si substitution do not show this behavior.

**Substitution transfer**

To investigate the impact of hydrogen substitution site transfer—from Mg to Si sites—on the physical properties of hydrous β-phase minerals, we performed a detailed series of simulations and analyses. This substitutional transition occurs in hydrous wadsleyite near the 410 km discontinuity, marking the boundary between the upper mantle and the MTZ. Although the two substitutional configurations exhibit trivial differences in volume and crystal structure, their contrasting bonding environments suggest potential variations in physical properties. Given the geophysical significance of such transitions, we focused our analysis on key properties including seismic wave velocities and electrical conductivities, which are critical for interpreting observations of Earth's deep interior.

We conducted a series of machine learning molecular dynamics (MLMD) simulations on α-Si, β-Mg and β-Si under high pressures and high temperature conditions. Simulations were performed with seismic velocities derived from stress–strain relationships obtained through simulations of distorted supercells. The high P–T compressional ($V_P$) and shear ($V_S$) wave velocities of these structures can be deduced from the elastic constants. As results, the $V_P$ and $V_S$ of β-Mg and β-Si at conditions correspond to MTZ are shown in Fig. 2a, which are consistent with the previous results[40-42,42]. Importantly, $V_P$ and $V_S$ values exhibits continuity across the subsitition, indicating that the Mg-to-Si site transfer does not significantly affect seismic wave propagation in hydrous wadsleyite. However, both $V_P$ and $V_S$ show a noticeable decrease upon entering high-temperature superionic states. This reduction is attributed to the onset of atomic diffusion, which softens the elastic moduli and consequently reduces wave velocities[43]. Specifically, we found a ~1 km/s drop in $V_P$ and a ~0.5 km/s reduction in $V_S$ as temperature increases from 1500 K to 2000 K.

The electrical conductivities are predicted from the MLMD simulations results using the Nernst–Einstein equation. Fig. 2b shows the electrical conductivities of β-Mg and β-Si under various temperatures at 14 GPa, which is the conditions that at the boundary between the upper mantle and MTZ (~410 km), coincident with the occurrence of substitution transfer. Our results indicated that electrical conductivity decreases by ~ 1.6 times as hydrogen substitution shifts from Mg into Si sites, accompanied by an increase in water content from 0.8 wt% to 1.6 wt%. This finding challenges the conventional expectation that higher water content directly enhances conductivity.

Instead, our results highlight that the specific nature of hydrogen substitution plays a critical role in governing conductivity The reduction in conductivity across the substitution transition provides key insight into the complex interplay between structural chemistry and geophysical observables in MTZ.

**Double Superionic behavior**

To investigate the dynamical properties and diffusion behavior of hydrous magnesium silicates under mantle-relevant conditions, we employed machine learning molecular dynamics (MLMD) simulations. These simulations utilize machine-learned force field (MLFF) potentials, which were trained on ab initio molecular dynamics (AIMD) data across a range of pressures and temperatures. The MLFF approach enables efficient simulations over long timescales and large supercells, overcoming the limitations of traditional AIMD, which is computationally expensive—especially when modeling low hydrogen concentrations. This approach is essential for our study, as hydrogen diffusion events are rare and typically inaccessible within the short timescales of conventional AIMD simulations. To capture these diffusion phenomena, we carried out MLMD simulations on supercells containing approximately 1000 atoms, with total simulation times of up to 300 picoseconds.

Our simulations reveal clear evidence of superionic behavior in α-Si, β-Mg, and β-Si at elevated temperatures. Fig. 3a–c and 3d–f show, respectively, the averaged mean squared displacements (MSDs) and atomic trajectories from MLMD simulations in the NVT ensemble. At low temperatures, atoms in all structures exhibit harmonic vibrations near their equilibrium positions, with MSD slopes remaining zero ($D_{H,O,Mg,Si}=0$), characteristic of solid-state behavior. As temperature increases, hydrogen ions begin to escape their original positions and diffuse freely through the relatively rigid sublattice formed by the other atoms. This results in non-zero MSD slopes for hydrogen ($D_H>0$, $D_{O,Mg,Si}=0$). This indicates that these hydrous minerals enter into a superionic state. Such superionic behavior is commonly observed in hydrogen-rich compounds under extreme conditions and is of broad interest in both materials science and planetary science applications [44-55].

Remarkably, we discovered the presence of *double superionicity* in hydrous wadsleyite phases (β-Mg and β-Si). Upon entering the hydrogen superionic regime (designated as superionic-I), a second transition occurs at temperatures exceeding ~2000 K, where magnesium ions also begin to diffuse ($D_{H,Mg}>0$, $D_{O,Si}=0$), marking the emergence of a superionic-II phase. This behavior is illustrated in Fig. 3g and 3h, where hydrogen diffusion is prominent and Mg shows a subtle but noticeable increase in MSD.

The atomic framework, largely maintained by oxygen and silicon atoms, provides a stable lattice through which both $H^+$ and $Mg^{2+}$ can migrate. Fig. 3i and 3j show representative atomic trajectories from MLMD simulations highlighting this double superionic behavior. Notably, $Mg^{2+}$ diffusion predominantly occurs near cation vacancies, which appear to act as preferential pathways or "tunnels" for magnesium migration. Such slow diffusion would likely be overlooked without the long-timescale simulations enabled by the MLMD method—underscoring its critical advantage.

In addition, we found our predicted hydrous ringwoodite (γ-Mg and γ-Si) possess the double superionicity as well (Fig. S3), even though they are not thermodynamically stable up to 100 GPa. However, this suggests that the behavior could be common and not unique in hydrous rocks under terrestrial conditions. The magnesium diffusion is considered to be one of the sources of electronic conductivities on anhydrous magnesium silicates[56], while few people have observed the coexisting diffusive hydrogen and magnesium phase in computational simulations. Kyla et al. reported the observation of double superionic states on icy H-C-N-O compounds using computer simulations under interior conditions of ice giants[57]. This study indicates that the double superionicity may exist in many hydrous terrestrial rocks under extreme conditions.

We constructed several systematic P-T phase diagrams (Fig. 4a-c) of α-Si, β-Mg and β-Si under Earth mantle conditions in order to obtain comprehensive understandings on their dynamical properties. The phase diagrams describe that all these hydrous minerals have distinct solid, superionic and liquid regions. α-Si has one hydrogen superionic state region and β-Mg and β-Si have two superionic state regions, including a hydrogen superionic state (superionic-I) and a double superionic state (superionic-II). These phase diagrams can help us to verify the specific states of these hydrous magnesium silicates in certain pressure and temperature conditions. It is noteworthy that the melting points of hydrous wadsleyite phases (β-Mg and β-Si) are about 1000 K higher than hydrous forsterite phases (α-Si). This is because of the existence of double superionic states in β-Mg and β-Si, which could cause the melting hysteresis in these hydrous wadsleyite phases and increase their melting temperatures. This discovery offers novel perspectives for the study of investigating the melting properties on hydrous magnesium silicates.

Finally, we calculated diffusion coefficients of $H^+$ and $Mg^{2+}$ under various pressures and temperatures conditions across all phases (Fig. 4d-f). The results illustrate that the diffusion coefficients of $H^+$ and $Mg^{2+}$ both increase under elevating temperatures, which indicate the transitions from solid into superionic states and eventually into liquid. Nevertheless, the diffusion coefficients of $Mg^{2+}$ are about an order of magnitude lower than those of $H^+$ in double superionic phases, which means $Mg^{2+}$ diffuse slower than $H^+$. Specifically, $H^+$ diffusion ranges from 0 to ~2 × 10⁻⁹ m²/s in α-Si, and from 0 to ~15 × 10⁻⁹ m²/s in β-Mg and β-Si across their respective superionic regimes. In comparison, $Mg^{2+}$ diffusion in the double superionic phases reaches up to ~1 × 10⁻⁹ m²/s.

**Discussion**

In this work, we identified a hydrogen substitution site transfer in hydrous wadsleyite occurring near ~410 km depth. This transition, from Mg-site to Si-site substitution, is accompanied by the formation of $SiO_2$, a major component of mid-ocean ridge basalt (MORB). This substitution results in a 1.6 fold decrease in the electrical conductivity.

This finding runs counter to the widely held assumption that higher water content leads to increased conductivity. Instead, our results demonstrate that the *substitution site itself* plays a non-negligible role in determining the electrical properties of hydrous minerals.

Given this substitution-induced effect, conductivity-based estimates of water content in the MTZ may require significant revision. When comparing structures with and without substitution effects but with the same measured conductivity, the inferred water content deviates by approximately 1 wt%. Specifically, our Si-substituted β-wadsleyite indicates a water content of ~1.6 wt%, whereas prior models estimates that neglect substitution effects report ~0.6-0.8 wt% for equivalent conductivity[16,36,37]. These results imply that the MTZ may contain more water than previously estimated, and they underscore the importance of accounting for substitution mechanisms in geophysical models.

Notably, this substitution transfer is *not* detectable through seismic velocity measurements, as both compressional and shear wave velocities remain continuous across the transition. This limits the utility of seismic data in detecting such substitutional transformations. Future work should investigate whether substitution transfer also affects other physical properties, and whether similar mechanisms are present in other hydrous minerals. We anticipate that further studies will enable construction of a comprehensive map relating substitution types to water content and physical observables.

In our study, we also predicted the presence of double superionic behaviors in hydrous magnesium silicates under extreme conditions. It suggests that such dual-ion mobility could be a more widespread phenomena among planetary materials. Experimental investigations have previously reported magnesium diffusion and Mg–Fe interdiffusion in magnesium silicate minerals[58,59], which supports the plausibility of cation mobility under mantle-like environments. We expect that the discover of the double superionicity in hydrous wadsleyite and ringwoodite can be validated in future experiments[60].

The double superionicity in hydrous $Mg_2SiO_4$ may play an important role in the evolution of super-Earths. The deep interiors of super-Earths have been discovered as dominant water reservoirs[61, 62], which provides favorable conditions for the formation of hydrous minerals including $Mg_2SiO_4$. At high temperatures, the double superionicity in these hydrous phases could increase the electrical conductivitiy and further affect the generation of magnetic dynamos, which is an interesting hypothesis besides of melting assumptions[63, 64]. Furthermore, the emergence of two superionic phases could lead to the additional layers inside these planets.

**Conclusion**

In summary, we have identified three thermodynamically stable hydrous magnesium silicates — α-Si, β-Mg, and β-Si — under Earth's mantle conditions through crystal structure prediction and first-principles calculations. All three phases exhibit hydrogen superionic behavior at elevated temperatures, and notably, hydrous wadsleyite phases

(β-Mg and β-Si) display *double superionicity*, characterized by the simultaneous diffusion of H$^+$ and Mg$^{2+}$ ions.

Importantly, there is a substitution transition in hydrous wadsleyite from Mg-substitution to Si-substitution at MTZ's conditions. The substitution transfer leads to electrical conductivity decreasing, which further influences the estimated water content in MTZ. This substitution transfer leads to a measurable decrease in electrical conductivity, despite an increase in water content from ~0.8 wt% to ~1.6 wt%. As a result, previously reported conductivity-based estimates of water content in MTZ — which did not account for substitution effects — may underestimate the true water budget by as much as ~1 wt%.

These findings suggest that the MTZ could serve as a significantly larger water reservoir than previously assumed. The substitution effect must therefore be considered in future models of mantle hydration and global water cycling. More broadly, our results enhance current understanding of the deep Earth's physical and chemical behavior and offer new perspectives on the structure, dynamics, and magnetic potential of super-Earth exoplanets. This work establishes a foundation for further investigation into superionic transport phenomena and substitution mechanisms in hydrous minerals under extreme conditions.

## Methods

### *Ab initio* calculations

*Ab initio* calculations are performed by Vienna *ab initio* simulation package (VASP) package[65] with the projector augmented wave (PAW) method[66]. We used the generalized gradient approximation (GGA) of Perdew-Burke-Ernzerhof (PBE)[67] for the exchange-correlation functional with the cut-off energy of 1000 eV. We treat 3s$^2$, 3s$^2$3p$^2$, 2s$^2$2p$^4$ and 1s$^1$ electrons as valence electrons for Mg, Si, O and H atoms, respectively. Monkhort-Pack *k*-point mesh[68] with 0.03·(2π/Å) and 0.04·(2π/Å) mesh resolution was respectively used for the structural optimization and phonon calculations. The phonon, zero-point energy (ZPE) calculations and quasi-harmonic approximations (QHA) calculations were performed by PHONOPY code[69]. The *ab initio* molecular dynamics (AIMD) simulations were conducted with NVT and NPT ensembles using the Langevin thermostat[70 错误!未找到引用源。] implemented by VASP. We used 800 eV for plane-wave energy cutoff and gamma point for k-point sampling, with total simulation times of 10 ps and timesteps of 0.5 fs. The supercells contain 115 atoms, 113 atoms and 115 atoms for α-Si, β-Mg and β-Si, respectively. Structure predictions were carried by AIRSS[71].

### Machine learning molecular dynamics (MLMD)

The machine learning force field (MLFF) potentials were constructed by the Deep Potential for Molecular Dynamics (DeePMD)[72, 73], in which a deep neural network is trained to learn and generate potentials using atomic coordinates, energies, and forces from AIMD simulations data. The machine learning molecular dynamics (MLMD) simulations were carried out using the LAMMPS code[74] with the force field potentials generated by DeePMD, employing periodic boundary conditions and a time step of 1 fs. The MLMD simulations comprised the supercells containing 920 atoms, 904 atoms and 920 atoms respectively for α-Si, β-Mg and β-Si under various pressures and temperatures for long-time scale of 300 ps in NVT and NPT ensembles.

**Electrical conductivity**

We utilized Nernst–Einstein equation ($\sigma=DNq^2/k_BT$) to calculate electrical conductivity. In the equation, q is the carrier electric charge, D is the carrier diffusion coefficient, N is the carrier density, and T is the temperature. All these values can be extracted from the AIMD or MLMD trajectories with NVT and NPT ensembles.

**Seismic velocities**

The seismic velocities were obtained by a series of processing approaches on the results of molecular dynamics simulations. The elastic constants $C_{ij}$ (i, j=1, 2, 3, 4, 5, 6) were calculated by distorting the equilibrium structure, conducting molecular simulations on the distorted structures in NVT ensembles and solving the strain–stress relations. The strain–stress relation is called Hooke's law, expressed as following:

$$\begin{pmatrix}\sigma_1\\\sigma_2\\\sigma_3\\\sigma_4\\\sigma_5\\\sigma_6\end{pmatrix}=\begin{pmatrix}C_{11}&C_{12}&C_{13}&C_{14}&C_{15}&C_{16}\\&C_{22}&C_{23}&C_{24}&C_{25}&C_{26}\\&&C_{33}&C_{34}&C_{35}&C_{36}\\&&&C_{44}&C_{45}&C_{46}\\&&&&C_{55}&C_{56}\\&&&&&C_{66}\end{pmatrix}\begin{pmatrix}\varepsilon_1\\\varepsilon_2\\\varepsilon_3\\\varepsilon_4\\\varepsilon_5\\\varepsilon_6\end{pmatrix}$$

Where σ refers to the stress tensor and ε refers to the strain tensor.

Then the bulk modulus (*B*) and shear modulus (*G*) were determined using the Voigt-Ruess-Hill scheme and we can further evaluate the primary wave velocity (*V<sub>P</sub>*) and shear wave velocity (*V<sub>S</sub>*):

$$V_P=\sqrt{\frac{B+\frac{4}{3}G}{\rho}},\ V_S=\sqrt{\frac{G}{\rho}}$$

**Acknowledgments**

We thank A. B. Belonoshko for the fruitful discussions, **Funding:** This work was supported by the National Natural Science Foundation of China (11774015, U2230401). D. Y. K. also acknowledges the support from Shanghai Science and Technology Committee, China (No. 22JC1410300) and Shanghai Key Laboratory of Material Frontiers Research in Extreme Environments, China (No. 22dz2260800).

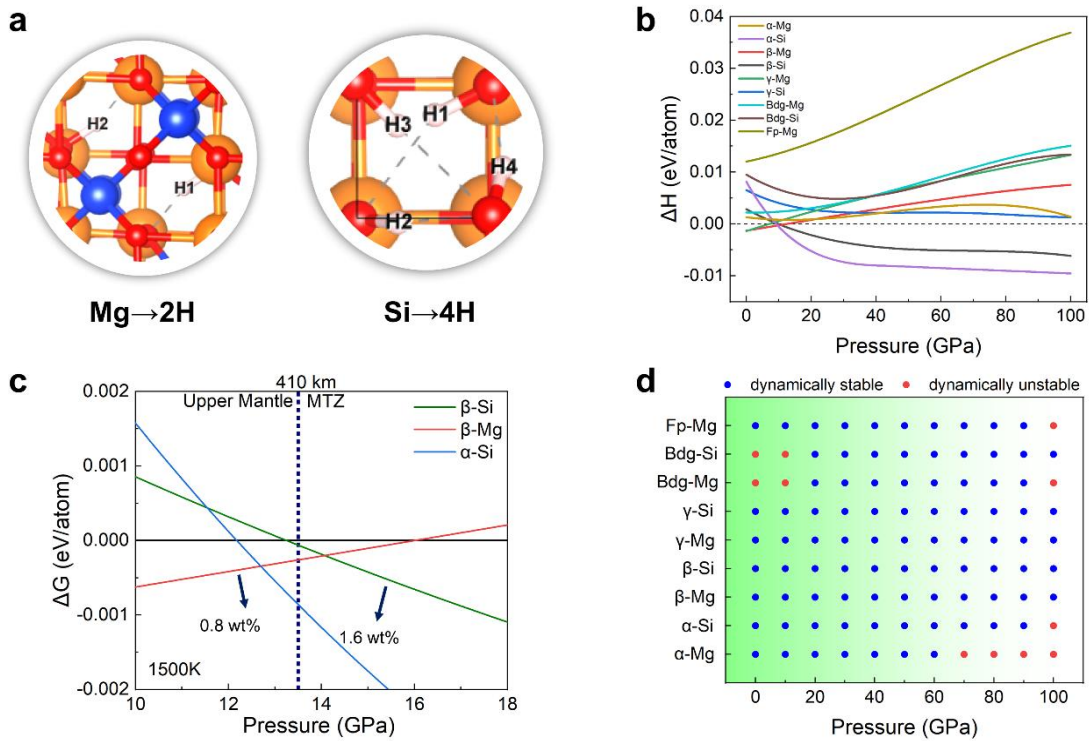

**Fig. 1. Stabilities of predicted hydrous minerals. a** The illustrations of hydrogen substitution on Mg site and Si site. **b** The formation enthalpies of all the predicted hydrous minerals included zero-point energy (ZPE) under 0-100 GPa. **c** Gibbs free energies of α-Si, β-Mg and β-Si by quasi-harmonic approximation (QHA). **d** Dynamically stable regions of predicted hydrous mineral under 0-100 GPa.

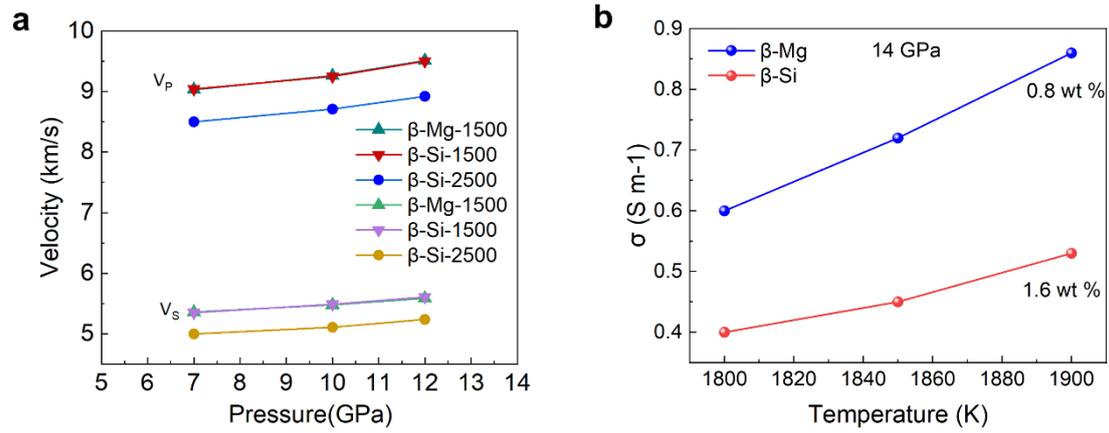

**Fig. 2 Physical properties during substitution transfer. a** Seismic velocities including compressional ($V_P$) and shear ($V_S$) wave velocities under transition zone conditions. **b** Electrical conductivities of β-Mg and β-Si under various temperatures at 14 GPa.

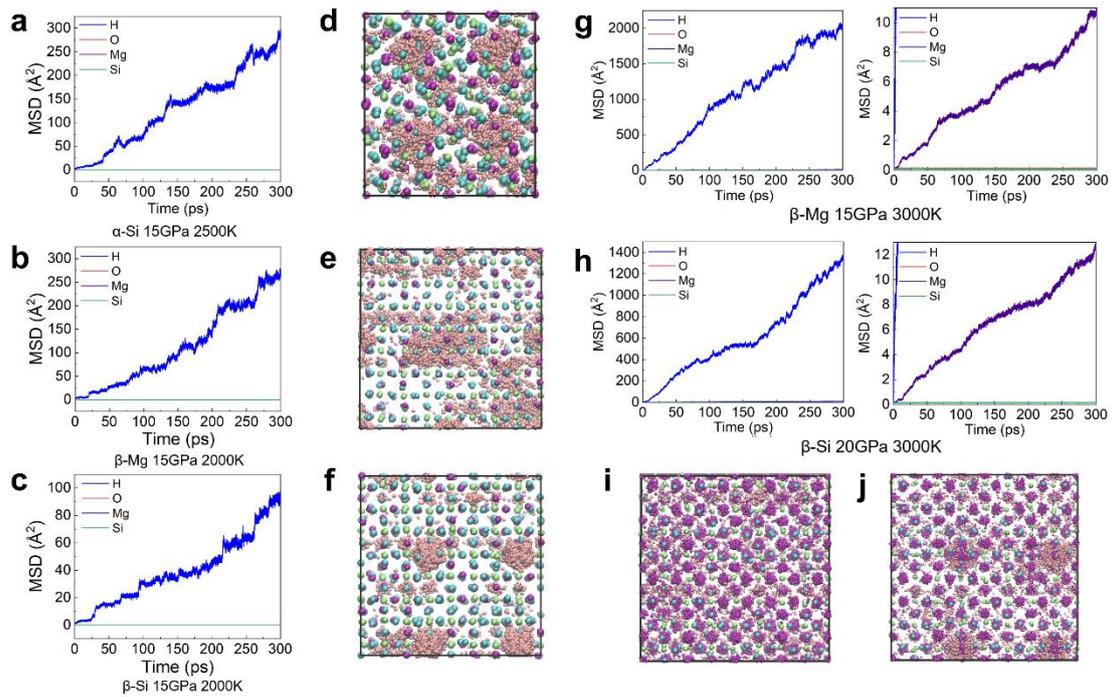

**Fig. 3. Dynamical behaviors of hydrous magnesium silicates at transition zone conditions. a-c** The mean square displacement (MSD) of β-Mg and β-Si under hydrogen superionic conditions. **d-f** The corresponding snapshots of MD trajectories in hydrogen superionic states. **g,h** The mean square displacement (MSD) of β-Mg and β-Si under double superionic conditions. **i,j** The corresponding snapshots of MD trajectories in double superionic states. In the snapshots of MD trajectories, the pink, blue, purple and green spheres represent H, O, Mg and Si atoms respectively.

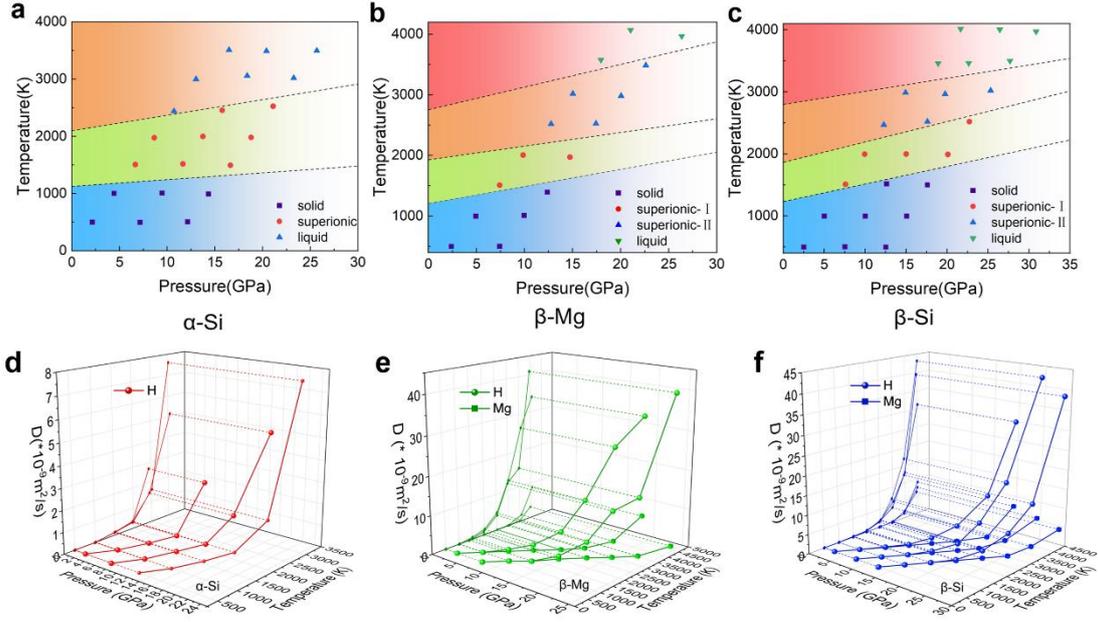

**Fig. 4. The phase diagram and diffusion coefficients of α-Si (a, d), β-Mg (b, e) and β-Si (c, f).** In α-Si phase diagram, The purple squares, red circles and blue triangle represent solid states, superionic states and liquid states, respectively. In β-Mg (b) and β-Si (c) phase diagrams, The purple squares, red circles, blue triangle and green triangles represent solid states, hydrogen superionic states, double superionic states and liquid states, respectively. The diffusion coefficients of hydrogen in α-Si (d), β-Mg (e) and β-Si (f) were represented by red, green and blue circles, respectively. The diffusion coefficients of oxygen in β-Mg (e) and β-Si (f) were represented by green and blue circles, respectively.